\RequirePackage{etoolbox}
\csdef{input@path}%
{%
 {sty/},
 {img/},
}
\documentclass[p04]{elsevierbook}

\input{mysymbol.sty}
\usepackage{graphicx}
\usepackage{bbm}
\usepackage{amsmath,amssymb,amsfonts}
\usepackage{url}

\newtheorem{mytheorem}{\bf Theorem}

\newtheorem{myproposition}{\bf Proposition}

\usepackage{natbib}
\begin{document}

\Frontmatter

\Mainmatter
  \begin{frontmatter}

\chapter{Topology Identification and Inference over Graphs}\label{chap1}


\begin{aug}
\author[addressrefs={ad1}]%
{%
\fnm{Gonzalo} \snm{Mateos}%
}%
\author[addressrefs={ad2}]%
{%
\fnm{Yanning} \snm{Shen}%
}%
\author[addressrefs={ad3}]%
{%
\fnm{Georgios B.} \snm{Giannakis}%
}%
\author[addressrefs={ad4}]%
{%
\fnm{Ananthram} \snm{Swami}%
}%
\address[id=ad1]%
{%
University of Rochester, Dept. of Electrical and Computer Engineering, Rochester, NY, United States.
}%
\address[id=ad2]%
{%
University of California Irvine, Dept. of Electrical Engineering and Computer Science, Irvine, CA, United States.
}%
\address[id=ad3]%
{%
University of Minnesota, Dept. of Electrical and Computer Engineering, Minneapolis, MN, United States.
}%
\address[id=ad4]%
{%
DEVCOM Army Research Lab.,
Adelphi, MD, United States.
}%
\end{aug}

\minitoc

\begin{abstract}
Topology identification and inference of processes evolving over graphs arise in timely applications involving brain, transportation, financial, power, as well as social and information networks. This chapter provides an overview of graph topology identification and statistical inference methods for multidimensional relational data. Approaches for undirected links connecting graph nodes are outlined, going all the way from correlation metrics to covariance selection, and revealing ties with smooth signal priors. To account for directional (possibly causal) relations among nodal variables and address the limitations of linear time-invariant models in handling dynamic as well as nonlinear dependencies, a principled framework is surveyed to capture these complexities through judiciously selected kernels from a prescribed dictionary. Generalizations are also described  via structural equations and vector autoregressions that can exploit attributes such as low rank, sparsity, acyclicity, and smoothness to model dynamic processes over possibly time-evolving topologies. It is argued that this approach supports both batch and online learning algorithms with convergence rate guarantees, is amenable to tensor (that is, multi-way array) formulations as well as decompositions that are well-suited for multidimensional network data, and can seamlessly leverage high-order statistical information.
\end{abstract}

\begin{keywords}
\kwd{Network topology identification}
\kwd{Graph-aware classification}
\kwd{Inference of network processes}
\end{keywords}

\end{frontmatter}

\section{Introduction}\label{sec:introduction}

As modern interconnected systems grow in size and importance,  and become more complex and heterogeneous, the need arises to advance a holistic theory of networks. At the same time, the information deluge propelled by the advent of e.g., online social media, the Web, and genome sequencing technologies has rendered statistical learning and inference from relational (graph) data indispensable. A symbiosis between the pervasive interest in scientific enquiry at a system level and the ever-growing capabilities for high-throughput data collection has only fueled this trend. Making sense of multidimensional datasets from a network-centric perspective will thus be a key enabler towards gaining new insights across science and engineering areas. 

The first step towards understanding network structure is identification of the underlying graph topology. Accordingly, a \emph{central question} is how to use nodal measurements to estimate latent graph connectivity, or, a judicious network model of said multidimensional data to facilitate efficient signal representation, visualization, prediction, (nonlinear) dimensionality reduction, and clustering. This chapter provides an overview of graph topology identification and statistical inference methods for multidimensional relational data; see also~\cite{giannakis18,dong2019learning,mateos19,kolaczyk2009book} for comprehensive tutorial treatments that inform (and complement) the material covered here.\vspace{2pt}

\noindent \textbf{Graph theoretic preliminaries and problem statement.} Consider a possibly weighted and directed graph $\ccalG(\ccalV,\ccalE,\bbW)$, where $\ccalV:=\{1,\ldots,N\}$ is the set of $N$ vertices or nodes, edges $\ccalE\subseteq \ccalV\times \ccalV$ comprise (ordered) pairs of vertices, and the generally \emph{unknown} adjacency matrix $\bbW\in\reals^{N\times N}$ has nonzero edge weights $W_{ij}\neq 0$ if a directed edge $(i,j)\in\ccalE$ is present from node $j$ to $i$. If $\ccalG$ is undirected then $W_{ij}\equiv W_{ji}$ so $\bbW$ is symmetric, and in the unweighted case $\bbW\in\{0,1\}^{N\times N}$ collects binary edge indicators $W_{ij}=\ind{(i,j)\in\ccalE}$. Suppose that the
network graph represents an abstraction of a complex system with measurable signals $y_{it}$ at nodes $i\in\ccalV$ and time $t=1,\ldots,T$, which we collect in $N \times 1$ network-wide vectors $\bby_t :=[y_{1t},\ldots,y_{Nt}]^\top$.
Vertex-indexed processes can model economic activity observed over a graph of production flows
among industrial sectors, or neural activity signals supported on brain connectomes, just to name a couple examples. We deal with the following problem.\vspace{2pt} 

\noindent\fbox{\parbox{0.99\textwidth}{\textbf{Topology identification.} Given a set $\ccalY:=\{\bby_t\}_{t=1}^T$ of nodal observations supported on the unknown graph $\ccalG(\ccalV,\ccalE, \bbW)$, the goal is to identify the topology encoded in the entries of the adjacency matrix $\bbW$ that is optimal in some sense. The optimality criterion is usually dictated by the adopted network-dependent model for the data in $\ccalY$, possibly augmented by priors motivated by physical characteristics of $\ccalG$, to effect statistical regularization, or else to favor more interpretable graphs.}}\vspace{2pt}

This is admittedly a fairly general and loose formulation that will be narrowed down in subsequent sections as we elaborate on various criteria stemming from different models binding the (statistical) signal properties to the graph topology.  Indeed, it is clear that one must assume \emph{some} relation between the signals and the unknown underlying graph, since otherwise the \emph{topology identification} exercise would be hopeless. This relation will be henceforth effected by statistical generative priors (Sections~\ref{ssec:correlations} and~\ref{ssec:graphical_lasso}), by properties of the signals with respect to the underlying graph such as smoothness (Section~\ref{ssec:smooth}), by linear structural equations and vector autoregressions (Section \ref{ssec:digraphs_causal}), or through judiciously selected kernels from a prescribed dictionary to account for directional (possibly causal) as well as nonlinear dependencies among nodal variables (Section \ref{ssec:nonlinear}). Structure identification methods for dynamic networks and multilayer graphs are the subject of Section \ref{ssec:dynamic_tensor}. Having acquired (or knowing a priori) the topology of a graph provides statistical information about relationships among nodes, and can thus be beneficial for \emph{inference of processes evolving over networks}. In this direction, we will also discuss task-aware graph identification algorithms for signal classification (Section \ref{ssec:discriminative_graph_learning}) and imputation (Section \ref{ssec:joint_topo_signal}).\vspace{2pt}

\noindent \textbf{Notational conventions.} Entries of a matrix $\bbX$ and a (column) vector $\bbx$ are denoted by $X_{ij}$ and $x_i$, respectively. For tensors, we adopt the notation $\underline{\bbX}$. Sets are represented by calligraphic capital letters. Symbols $\bbzero$ and $\bbone$ refer to the all-zero and all-one vectors; while $\bbI$ denotes the identity matrix. For a vector $\bbx$, $\diag(\mathbf{x})$ is a diagonal matrix whose $i^{th}$ diagonal entry is $x_i$. The operators $\circ$, $(\cdot)^\top$, $\text{trace}(\cdot)$, and $\text{vec}(\cdot)$ stand for Hadamard (element-wise) product, transposition, matrix trace, and vectorization, respectively. The indicator function $\ind{\textsc{cond}}=1$ if \textsc{cond} holds, and is $0$ otherwise. Lastly, $\| \bbX \|_p$ denotes the $\ell_p$ norm of $\text{vec}(\bbX)$ and $\| \bbX \|_F$ refers to the Frobenius norm. 

\section{A tale of signal correlations and smoothness}\label{sec:undirected}

Long-standing correlation-based network topology identification approaches are the main subject of our initial exposition in this section. Through a natural and didactic progression, we then show how these statistical methods connect to a recent class of approaches that aim to construct a graph on which
network data admit certain ``regularity.'' Optimization and downstream task-aware graph learning advances are discussed. The initial focus is on identifying symmetric, linear, and static pairwise relations between network elements, which we markedly expand on in Section \ref{sec:dnd}.

\subsection{Correlation networks}\label{ssec:correlations}

Arguably, the most widely adopted linear measure of similarity between nodal random variables $y_i$ and $y_j$ is the Pearson correlation coefficient $\rho_{ij}$ that is given by
\begin{equation}\label{E:corr_coeff}
	\rho_{ij}:=\frac{\textrm{cov}(y_i,y_j)}{\sqrt{\textrm{var}(y_i)\textrm{var}(y_j)}},
\end{equation}
and can be obtained from entries $\sigma_{ij}:=\textrm{cov}(y_i,y_j)$ in the covariance matrix $\bbSigma:=\E{(\bby-\bbmu)(\bby-\bbmu)^{\top}}$ of the random (graph) signal $\bby=[y_1,\ldots,y_N]^{\top}$, with mean vector $\bbmu:=\E{\bby}$. Given this choice, it is natural to define the \emph{correlation network} $\ccalG(\ccalV,\ccalE, \bbW)$ with vertices $\ccalV:=\{1,\ldots,N\}$ and edge set $\ccalE:=\{(i,j)\in\ccalV\times \ccalV:\rho_{ij}\neq 0\}$. There is some latitude on the definition of the weights. To directly capture the correlation strength between $y_i$ and $y_j$ one can set $W_{ij}=\rho_{ij}$ or its un-normalized variant $W_{ij}=\textrm{cov}(y_i,y_j)$; alternatively, the choice $W_{ij}=\ind{\rho_{ij}\neq 0}$ gives an unweighted (binary) graph consistent with $\ccalE$. Regardless of these choices, what is important here is that the definition of $\ccalE$ dictates that the task of identifying the topology of $\ccalG$ becomes one of inferring the subset of nonzero correlations.  

To that end, given $T$ independent realizations $\ccalY:=\{\bby_t\}_{t=1}^T$ of $\bby$ one forms empirical correlations $\hat{\rho}_{ij}$ by replacing the ensemble covariances in \eqref{E:corr_coeff}, with the entries $\hat{\sigma}_{ij}$ of the unbiased sample covariance matrix estimate $\hbSigma$. One could then manually fix a threshold and assign edges to the corresponding largest values $|\hat{\rho}_{ij}|$, say, given a target density. Instead, a more principled approach is to test the hypotheses
\begin{equation}\label{E:corr_tests}
	H_0:\rho_{ij}=0\quad \textrm{versus}\quad  H_1:\rho_{ij}\neq 0,
\end{equation}
for each of the ${N \choose 2}=N(N-1)/2$ candidate edges in $\ccalG$, i.e., the number of unordered pairs in $\ccalV \times \ccalV$. While $|\hat{\rho}_{ij}|$ would appear to be the go-to test statistic, a more convenient choice is the Fisher score $z_{ij}:=\tanh^{-1}(\hat{\rho}_{ij})$. The reason is that  under $H_0$ one (approximately) has $z_{ij}\sim \textrm{Normal}(0,\frac{1}{T-3})$; see~\cite[p. 210]{kolaczyk2009book} for further details and the justification based on asymptotic-theory arguments. This simple form of the null distribution facilitates computation of $p$-values, or, the selection of a threshold that guarantees a prescribed significance level per edgewise test. \vspace{2pt}

\noindent \textbf{The multiple testing problem.} Such individual test control procedures might be ineffective for moderate to large-sized graphs, since the total number of simultaneous tests to be conducted scales as $O(N^2)$. Leaving aside potential computational challenges, \emph{multiple testing} issues must be addressed~\cite[Ch. 15]{efron2016book}. Otherwise, say for an empty graph with $\ccalE=\emptyset$, a constant false alarm rate $P_{FA}$ per edge will yield on average $O(N^2P_{FA})$ spurious edges, which can be prohibitive if $N$ is large. A common workaround is to control the false discovery rate (FDR) defined as
\begin{equation}\label{E:FDR}
	\textrm{FDR}:=\E{\frac{R_{f}}{R}\given R>0}\Pr{R>0}
\end{equation}
where $R$ denotes the number of rejections among all $O(N^2)$ edgewise tests conducted, and $R_f$ stands for the number of false rejections (i.e., false edge discoveries). Let $p_{(1)}\leq p_{(2)}\leq\ldots\leq p_{(|\ccalV\times\ccalV|)}$ be the ordered $p$-values for all tests. Then a prescribed level $\textrm{FDR}\leq q$ can be guaranteed by following the Benjamini-Hochberg FDR control procedure, which declares edges for all tests $i$ such that $p_{(i)}\leq (2i/N(N-1))q$; see e.g.,~\cite[Ch. 15.2]{efron2016book}. The FDR guarantee is only valid for independent tests, an assumption that rarely holds in a topology identification setting. Hence, results and control levels should be interpreted with care; see also~\cite[p. 212]{kolaczyk2009book} for a discussion on FDR extensions when some level of dependency is present between tests.\vspace{2pt}  

\noindent \textbf{Controlling for confounders.} On the scope of correlation networks, apparently they can only capture linear and symmetric pairwise dependencies among vertex-indexed random variables.  A topology identification algorithm that exploits higher-order statistics of the data is presented in \cite{smithTN2022}; there the nodal variables correspond to cumulative path delays or packet drop rates. Most importantly, the measured correlations can be due to latent network effects rather than from strong direct influence among a pair of vertices. For instance, a suspected regulatory interaction among genes $(i,j)$ inferred from their highly correlated micro-array expression-level profiles, could be an artifact due to a third latent gene $k$ (a.k.a. a confounder) that is actually regulating the expression of both $i$ and $j$. If seeking a graph reflective of \emph{direct influence} among pairwise signal elements, clearly correlation networks may be undesirable. 

One can resolve such a confounding by instead considering \emph{partial} correlations  
\begin{equation}\label{E:par_corr_coeff}
	\rho_{ij|\ccalV\setminus \{ij\}}:=\frac{\textrm{cov}(y_i,y_j\given \ccalV\setminus \{ij\})}{\sqrt{\textrm{var}(y_i\given \ccalV\setminus \{ij\})\textrm{var}(y_j\given \ccalV\setminus \{ij\})}},
\end{equation}
where $\ccalV\setminus \{ij\}$ symbolically denotes the collection of all $N-2$ random variables $\{y_k\}$ after excluding those indexed by nodes $i$ and $j$. A \emph{partial correlation network} can be defined in analogy to its (unconditional) correlation network counterpart, but with edge set $\ccalE:=\{(i,j)\in\ccalV\times \ccalV:\rho_{ij|\ccalV\setminus \{ij\}}\neq 0\}$. Again, inferring nonzero partial correlations from data 
$\ccalY:=\{\bby_t\}_{t=1}^T$ can be tackled via hypothesis testing. With minor twists, issues of selecting a test statistic and a tractable approximate null distribution, as well as dealing with the multiple testing problem, can all be addressed by following similar guidelines to those in the Pearson correlation case~\cite[Ch. 7.3.2]{kolaczyk2009book}.

\subsection{From covariance selection to graphical lasso}\label{ssec:graphical_lasso}

Suppose now that $\bby := [y_1,\ldots,y_N]^{\top}$ is a Gaussian random vector, which means that the vertex-indexed random variables are jointly Gaussian. Under such a distributional assumption, $\rho_{ij|\ccalV\setminus \{ij\}}=0$ is equivalent to $y_i$ and $y_j$ being conditionally \emph{independent} given all of the other variables in  $\ccalV\setminus \{ij\}$. Consequently, the partial correlation network with edges $\ccalE:=\{(i,j)\in\ccalV\times \ccalV:\rho_{ij|\ccalV\setminus \{ij\}}\neq 0\}$ specifies conditional (in)dependence relations among the entries of $\bby$, and is known as an undirected Gaussian graphical model or Gaussian Markov random field (GMRF).

A host of opportunities for inference of Gaussian graphical models emerge by recognizing that the partial correlation coefficients can be expressed as
\begin{equation}\label{E:par_corr_precision}
	\rho_{ij|\ccalV\setminus \{ij\}}=-\frac{\theta_{ij}}{\sqrt{\theta_{ii}\theta_{jj}}}
\end{equation}
where $\theta_{ij}$ is the $(i,j)$-th entry of the precision matrix $\bbTheta:=\bbSigma^{-1}$, the inverse of the covariance matrix $\bbSigma$ of $\bby$. The upshot of \eqref{E:par_corr_precision} is that it reveals a bijection between the set of nonzero partial correlations (the edges of $\ccalG$) and the support of the precision matrix $\bbTheta$. The graphical model selection task of identifying the conditional independence relations in $\ccalG$ given i.i.d. realizations $\ccalY:=\{\bby_t\}_{t=1}^T$ from a multivariate Gaussian distribution, is known as the \emph{covariance selection} problem. The term covariance selection was first coined in the 70's by Dempster, who explored the role of sparsity in estimating the entries of $\bbTheta$ via a recursive, likelihood-based thresholding procedure on the entries of $\hbTheta:=\hbSigma^{-1}$~\cite{dempster_cov_selec}. In addition to facing computational challenges, in modern high-dimensional regimes where $N\gg T$ the method breaks down since the sample covariance matrix $\hbSigma$ is rank deficient. Such a predicament calls for regularization, and next we briefly outline prominent graphical model selection approaches based on $\ell_1$-norm regularized global likelihoods for the Gaussian setting.\vspace{2pt} 

\noindent \textbf{Graphical lasso.} We henceforth assume zero-mean $\bby\sim\textrm{Normal}(\mathbf{0},\bbSigma)$, since the focus is on estimating the graph structure encoded in the entries of the precision matrix $\bbTheta=\bbSigma^{-1}$. 
%
%
The \emph{graphical lasso} regularizes the maximum likelihood estimator of the precision matrix with the sparsity-promoting $\ell_1$-norm of $\bbTheta$~\cite{yuanlin2007}, yielding 
\begin{equation}\label{E:glasso}
	\hbTheta\in\arg\max_{\bbTheta\succeq\mathbf{0}}\left\{\log\det\bbTheta -\textrm{trace}(\hbSigma \bbTheta)-\lambda\|\bbTheta\|_1\right\}.
\end{equation}
%
%
The constraint $\bbTheta\succeq\mathbf{0}$ enforces the matrix to be positive semidefinite (PSD), and $\hbSigma$ is the empirical covariance matrix obtained from the data in $\ccalY$. 
Although \eqref{E:glasso} is convex, the objective is non-smooth and has an unbounded constraint set. The complexity for off-the-shelf interior point methods adopted in~\cite{yuanlin2007} is $O(N^6)$, 
thus prohibitive for even modest-sized graphs. 
Efficient first-order cyclic block-coordinate descent algorithms were developed in~\cite{banerjee2008jlmr} and subsequently refined in~\cite{glasso2008}, which can comfortably tackle sparse problems with thousands of nodes in under a few minutes. In terms of performance guarantees for the recovery of a ground-truth precision matrix $\bbTheta_0$, the graphical lasso estimator \eqref{E:glasso} with ${\lambda=2\sqrt{T^{-1}\log N}}$ satisfies the operator norm bound ${\|\hbTheta-\bbTheta_0\|_2\leq \sqrt{d_{\max}^2T^{-1}\log N}}$ with high probability, where $d_{\max}$ denotes the maximum nodal degree in $\bbTheta_0$~\cite{ravikumar2011}. Support consistency has been also established provided that the number of samples scales as $T=\Omega(d_{\max}^2\log N)$; see~\cite{ravikumar2011} for details. \vspace{2pt}

\noindent\textbf{Learning Gaussian graphical models with Laplacian constraints.} A significant body of work~\cite{Lake10discoveringstructure,egilmez2017jstsp,mike_icassp17,yingneurips20,kumar2020jmlr} advocates estimating precision matrices subject to combinatorial graph Laplacian constraints [for undirected $\ccalG(\ccalV,\ccalE,\bbW)$, the graph Laplacian matrix is $\bbL=\textrm{diag}(\bbW\mathbf{1})-\bbW$]. The off-diagonal elements $L_{ij}:=-W_{ij}\leq 0$ of a Laplacian must be non-positive, so when $\bbTheta=\bbL$ the resulting GMRF is termed \emph{attractive}; see also~\cite{slawski2015mmatrices} for estimation of precision matrices under the constraint that all partial correlations are non-negative (often useful for interpretability). Moreover, the Laplacian matrix is always singular because $\bbL\mathbf{1}=\mathbf{0}$, which yields an \emph{improper} GMRF. A proper GMRF can be obtained via diagonal loading of the sought Laplacian, which motivates the following sparse precision matrix estimation problem~\cite{Lake10discoveringstructure}
\begin{align}\label{E:GMRF_Laplacian}
&{}\max_{\bbTheta,\gamma\geq 0}\left\{\log\det\bbTheta -\textrm{trace}(\hbSigma \bbTheta)-\lambda\|\bbTheta\|_1\right\}\\
&{}\textrm{subject to }\:\bbTheta = \bbL+\gamma\bbI,\: \bbL\mathbf{1}=\mathbf{0},\: L_{ij}\leq 0, \:i\neq j.\nonumber
\end{align}
Given a solution $\{\hbTheta,\hat{\gamma}\}$ of \eqref{E:GMRF_Laplacian}, a combinatorial Laplacian can be recovered as $\hbL:=\hbTheta-\hat{\gamma}\bbI$. There are various probabilistic interpretations of such a diagonal loading, e.g., one where $\gamma^{-1}$ corresponds to the variance of white Gaussian noise modeling isotropic signal fluctuations. In this context, a general optimization framework for estimating (possibly diagonally-dominant, generalized) Laplacian matrices was presented in~\cite{egilmez2017jstsp}. 
Recent advances in~\cite{kumar2020jmlr} contribute a comprehensive approach for structured (connected, multi-component, bipartite) graph learning via Laplacian spectral constraints along with efficient block majorization-minimization solvers for the resulting problems; see also~\cite{viniciusneurips21} for applications with heavy-tailed financial data.
    
Interestingly, when $\bbTheta=\bbL$ notice that minimization of the term
\begin{equation}\label{eq:tv}
\textrm{trace}(\hbSigma \bbL)\propto \sum_{t=1}^T\bby_t^{\top}\bbL\bby_t=\sum_{t=1}^T\text{TV}(\bby_t) 
\end{equation}
in \eqref{E:glasso}-\eqref{E:GMRF_Laplacian} favors graphs for which the observed signals are smooth. This follows because $\text{TV}(\bby):=\bby^{\top}\bbL\bby=\frac{1}{2}\sum_{i\neq j}W_{ij}(y_i-y_j)^2$ is a total variation (TV) measure (or Dirichlet energy) of signal $\bby$ with respect to $\ccalG$; see e.g.,~\cite{kalofolias16}. The ubiquity of smooth network data has been well documented, with examples spanning sensor measurements, protein function annotations, and product ratings. These empirical findings motivate adopting smoothness as the criterion to search for graphs on which measurements exhibit desirable parsimony or regularity; the subject dealt with next.

\subsection{Graph learning from observations of smooth signals}\label{ssec:smooth}

The aforementioned discussion suggests the following network topology identification problem. Given nodal observations $\ccalY:=\{\bby_t\}_{t=1}^T$, the goal is to learn an undirected graph $\ccalG(\ccalV,\ccalE, \bbW)$ such that the signals in $\ccalY$ are smooth on $\ccalG$.

To this end, consider the data matrix $\bbY:=[\bby_1,\ldots,\bby_T]\in \reals^{N\times T}$, whose columns $\{ \bby_t\}_{t=1}^T$ are the observations in $\ccalY$. The rows of $\bbY$, denoted by $\bar{\bby}_i^{\top}\in\reals^{1\times T}$, collect all $T$ measurements at vertex $i$. Next, form the nodal Euclidean distance matrix $\bbE\in\reals_{+}^{N\times N}$, where $E_{ij}:=\|\bar{\bby}_i-\bar{\bby}_j\|_2^2$, $i,j\in\ccalV$. Using these definitions, the aggregate signal smoothness measure over $\ccalY$ in \eqref{eq:tv} can be equivalently written as~\cite{kalofolias16}
\begin{equation}\label{E:smooth_sparse}
	\sum_{t=1}^T\textrm{TV}(\bby_t)=\textrm{trace}(\bbY^{\top}\bbL\bbY)=\frac{1}{2}\|\bbW\circ\bbE\|_{1}
\end{equation}
where $\circ$ denotes the element-wise (Hadamard) product. Identity \eqref{E:smooth_sparse} reveals that the smoothness minimization criterion has the following intuitive interpretation: When pairwise nodal distances in $\bbE$ are sampled from a \emph{smooth} manifold, the learnt topology $\bbW$ tends to be \emph{sparse}, preferentially choosing edges $(i,j)$ whose corresponding $E_{ij}$ are smaller [cf. the weighted $\ell_1$-norm in \eqref{E:smooth_sparse}]. Leveraging this neat link between signal smoothness and edge sparsity, a fairly general graph-learning framework was put forth in~\cite{kalofolias16}. The idea therein is to solve the following convex inverse problem
\begin{align}\label{eq:kalofolias}
	&{}\min_{\bbW}\left\{\|\bbW\circ\bbE\|_1-\alpha\bbone^{\top} \log \left( \bbW\bbone \right)+\frac{\beta}{2}\|\bbW\|_F^2\right\}\\ 
	&{}\textrm{subject to }\:\textrm{diag}(\bbW)=\mathbf{0},\: W_{ij}=W_{ji}\geq 0, \:i\neq j \nonumber
\end{align}
where $\alpha,\beta>0$ are tunable regularization parameters. The logarithmic barrier on the vertex degrees $\bbd=\bbW\bbone$ excludes the possibility of having (often undesirable) isolated vertices in the estimated graph. Through $\beta$, the Frobenius-norm penalty offers a handle on the graphs' edge sparsity level; the sparsest graph is obtained when $\beta=0$.\vspace{2pt}
	
\noindent \textbf{Variable splitting, the dual problem and its favorable properties.} Since $\diag(\bbW) = \mathbf{0}$ and $\bbW$ is symmetric, the optimization variables in \eqref{eq:kalofolias} are effectively the, say, upper-triangular elements $W_{ij}$, $j>i$. Thus, it suffices to retain only those entries in the vector $\bbw:=\textrm{vec}[\textrm{triu}[\bbW]] \in \reals_{+}^{N(N-1)/2}$, applying matrix vectorization and upper-triangular extraction operators. To enforce nonnegative edge weights, we penalize the cost with the indicator $\mathbb{I}_{\infty}\{\bbw\succeq\mathbf{0}\}=0$ if $\bbw\succeq \mathbf{0}$, else $\mathbb{I}_{\infty}\{\bbw\succeq \mathbf{0}\}=\infty$. This way, we equivalently reformulate \eqref{eq:kalofolias} as the unconstrained, non-differentiable problem
\begin{equation}\label{eq:kalofolias_vec}
	\min_{\bbw}\Big\{\underbrace{\mathbb{I}_{\infty}\{\bbw\succeq\mathbf{0}\} + 2\bbw^{\top}\bbe+\beta\|\bbw\|_2^2}_{:=f(\bbw)}-\underbrace{\alpha \bbone^{\top} \log \left( \bbS\bbw \right)}_{ :=-g(\bbS\bbw)}\Big\} 
\end{equation}
where $\bbe:=\textrm{vec}[\textrm{triu}[\bbE]] $ and $\bbS\in\{0,1\}^{N\times N(N-1)/2}$ maps edge weights to nodal degrees, that is, $\bbd=\bbS\bbw$. The non-smooth function $f(\bbw):=\mathbb{I}_{\infty}\{\bbw\succeq\mathbf{0}\} + 2\bbw^{\top}\bbe+\beta\|\bbw\|_2^2$ is strongly convex with strong convexity parameter $2\beta$, while $g(\bbw):=-\alpha \bbone^{\top} \log \left(\bbw \right)$ is a (strictly) convex function for all $\bbw \succ \mathbf{0}$. Under these properties of $f$ and $g$, the composite problem \eqref{eq:kalofolias_vec} has a unique optimal solution $\bbw^\star$; see e.g.,~\cite{beck2014} and~\cite{wang2021}.

The structure of \eqref{eq:kalofolias_vec} lends itself naturally to variable splitting, so we write 
\begin{equation}\label{eq:primal_split}
	\min_{\bbw,\bbd}\left\{f(\bbw)+g(\bbd)\right\},\quad  \textrm{ s. to }\bbd=\bbS\bbw.
\end{equation}
Attaching multipliers $\bblambda\in \reals^N$ to the constraints and maximizing the Lagrangian $\ccalL(\bbw,\bbd,\bblambda)=f(\bbw)+g(\bbd)-\langle\bblambda,\bbS\bbw-\bbd\rangle$ with respect to the primal variables $\{\bbw,\bbd\}$, one arrives at the (minimization form) dual problem
\begin{equation}\label{eq:dual}
	\min_{\bblambda}\left\{F(\bblambda)+G(\bblambda)\right\}
\end{equation}
where the so-termed Fenchel conjugates of $f$ (composed with $\bbS \bbw$) and $g$ are given by
\begin{equation}
	F(\bblambda):=\max_{\bbw}\big\{\langle \bbS^\top \bblambda,\bbw\rangle-f(\bbw)\big\}\quad \textrm{ and }\quad 
	G(\bblambda):=\max_{\bbd}\big\{\langle -\bblambda,\bbd\rangle-g(\bbd)\big\}.\label{eq:conjugate_F}
\end{equation}
Interestingly, the strong convexity of $f$ induces useful smoothness properties for $F$~\cite[Lemma 3.1]{beck2014}. One can show
$F(\bblambda)$ in \eqref{eq:conjugate_F} is continuously differentiable, and the gradient $\nabla F(\bblambda)$ is Lipschitz continuous with constant $L:=\frac{N-1}{\beta}$. This additional structure of \eqref{eq:dual} makes it feasible to apply accelerated proximal gradient (PG) algorithms~\cite{beck18} (such as FISTA~\cite{beck2009}), to solve the dual problem \eqref{eq:dual}.\vspace{2pt}	

\noindent \textbf{Fast dual-based graph learning.} The FISTA algorithm applied to the dual problem \eqref{eq:dual} yields the following iterations (initialized with $\bbomega_{1}=\bblambda_0\in \reals^N$ and $t_1=1$; henceforth, $k=1,2,\ldots$ denotes the iteration index)
\begin{align}
\bblambda_k={}&\textbf{prox}_{L^{-1}G}\left(\bbomega_{k} - L^{-1} \nabla F(\bbomega_{k}) \right)\label{eq:FISTA_prox}\\
t_{k+1}={}&1/2+\sqrt{1/4+t_{k}^2} \label{eq:FISTA_t}\\
\bbomega_{k+1}={}& \bblambda_k+\left(\frac{t_k-1}{t_{k+1}}\right)\left[ \bblambda_k- \bblambda_{k-1}\right]	\label{eq:FISTA_extrapolation}
\end{align}
where $\textbf{prox}_{h}(\cdot)$ denotes the proximal operator of a proper, lower semi-continuous convex function $h$. 
%
%
An adaptation of the result in~\cite[Lemma 3.2]{beck2014} -- stated as Proposition \ref{prop:iterations} below -- yields a graph learning algorithm with several attractive features. 
\begin{myproposition}{\normalfont\cite{saman2021spl}}\label{prop:iterations}
The dual variable update iteration in \eqref{eq:FISTA_prox} can be equivalently rewritten as $\bblambda_k=\bbomega_k-L^{-1}(\bbS\bar{\bbw}_k-\bbu_k)$, with
\begin{align} 	
\bar{\bbw}_k={}& \max\left(\mathbf{0},\frac{\bbS^\top\bbomega_k-2\bbe}{2\beta}\right)\label{eq:barw_update} \\ 
\bbu_k={}&\frac{\bbS\bar{\bbw}_k-L\bbomega_k  + \sqrt{(\bbS\bar{\bbw}_k-L\bbomega_k)^2 + 4\alpha L\bbone}}{2} \label{eq:u_update}
\end{align}	
where $\max(\cdot,\cdot)$ in \eqref{eq:barw_update} as well as both $(\cdot)^2$ and $\sqrt{(\cdot)}$ in \eqref{eq:u_update} are element-wise operations on their vector arguments.
\end{myproposition}
The updates in Proposition \ref{prop:iterations} are fully expressible in terms of parameters from the original graph learning problem, namely $N$, $\alpha$, $\beta$, $\bbS$ and the data in $\bbe$. This is to be contrasted with \eqref{eq:FISTA_prox}, which necessitates the conjugate functions $F$ and $G$. 
The overall complexity of $O(N^2)$ is on par with state-of-the-art linearized alternating direction method of multipliers (ADMM) algorithms~\cite{wang2021}, which have been shown to scale well to relatively large networks with $N$ in the order of thousands. 
For a given problem instance, there are no step-size parameters to tune here (on top of $\alpha$ and $\beta$) since we can explicitly compute the Lipschitz constant $L:=\frac{N-1}{\beta}$. On the other hand, the linearized ADMM algorithm in~\cite{wang2021} necessitates tuning two step-sizes and the penalty parameter defining the augmented Lagrangian. 

The distinct feature of the accelerated dual PG algorithm is that it comes with global convergence rate guarantees. Indeed, when $k\to\infty$ the iterates $\bblambda_k$ generated by the update rule in Proposition \ref{prop:iterations} provably approach a dual optimal solution $\bblambda^\star$ that minimizes $\varphi(\bblambda):=F(\bblambda)+G(\bblambda)$ in \eqref{eq:dual}; see e.g.,~\cite{beck2009}. 
%
%
%
%
The well-documented $O(1/k^2)$ global convergence rate of accelerated PG algorithms implies an $\ccalO(1/\sqrt{\epsilon})$ iteration complexity to return an $\epsilon$-optimal dual solution measured in terms of $\varphi$ values. Interestingly, leveraging~\cite[Theorem 4.1]{beck2014} one can show the primal sequence  $\hbw_k=\argmin_{\bbw}\ccalL(\bbw,\bbd,\bblambda_k)$
is globally convergent to $\bbw^\star$ at a rate of $O(1/k)$. 
%
%
\begin{mytheorem}{\normalfont\cite[Theorem 2]{saman2021spl}}\label{th:FISTA_rate_primal_var}
For all $k\geq 1$, the primal sequence 
\begin{equation}
\hbw_k=\argmax_\bbw\left\{\langle \bbS^\top \bblambda_k,\bbw\rangle-f(\bbw)\right\}=\max\left(\mathbf{0},\frac{\bbS^\top\bblambda_k-2\bbe}{2\beta}\right) \label{eq:primal_seq}
\end{equation}
defined in terms of dual iterates $\bblambda_{k}$ in Proposition \ref{prop:iterations}, satisfies
\begin{equation}\label{eq.th1}
\|\hbw_k-\bbw^\star\|_2\leq \frac{\sqrt{2(N-1)}\|\bblambda_0-\bblambda^\star\|_2}{\beta k}.
\end{equation}
\end{mytheorem}

\subsection{Discriminative graph learning for classification}\label{ssec:discriminative_graph_learning}

Aiming to tackle classification tasks involving network
data, we show how the general approach in Section \ref{ssec:smooth} can be leveraged to obtain \emph{discriminative} graph-based signal representations. Consider a dataset $\ccalY=\bigcup_{c=1} ^C \ccalY_c$ comprising labeled observations $\ccalY_c:=\{\bby_{t}^{(c)}\}_{t=1}^{T_c}$ from $C$ different classes. The signals in each class possess a very distinctive structure, namely they are assumed to be smooth with respect to unknown class-specific graphs $\ccalG_c =\left(\ccalV,\ccalE_c,\bbW_c \right), \; c=1,\dots,C$.  This notion is analogous to a multiple linear subspace model in which data of each class are assumed to be spanned by a few Laplacian eigenvectors of the class-conditional graphs;  namely, the low-frequency graph Fourier transform (GFT) modes. 
Given $\ccalY$ the goal is to learn the class-specific adjacency matrices $\bbW_c$ under signal smoothness priors, so that the GFT bases obtained as a byproduct can be subsequently used to classify unseen (and unlabeled) signals effectively via low-pass filtering.  

The approach in~\cite{saboksayr20} blends ingredients from the graph learning framework in~\cite{kalofolias16} [cf. \eqref{eq:kalofolias}] along with the discriminative graphical lasso estimator~\cite{kao17}. Indeed, the algorithm in Section \ref{ssec:smooth} optimizes a topology recovery objective under smoothness assumptions, but is otherwise agnostic to the performance of a potential downstream (say, classification) task the learnt graph may be integral to. Inspired by~\cite{kao17}, the problem dealt with here calls for graph representations that capture the underlying network topology (that is, the class structure), but at the same time are discriminative to boost classification performance. To this end, the idea in~\cite{saboksayr20} is to learn a graph $\ccalG_c$ per class by solving the following convex optimization problems [cf. \eqref{eq:kalofolias}]
\begin{equation}\label{E:discriminative}
\min_{\bbW_c\in \ccalW}\:\left\{ \| \bbW_c \circ \bbE_c\|_{1} - \alpha\bbone^{\top}\log\left( \bbW_c \bbone \right) \\+ \beta \|\bbW_c\|_{F}^2 - \gamma\sum_{k\neq c}^{C} \| \bbW_c \circ \bbE_k\|_{1}\right\},
\end{equation}
where $\bbW_c$ is the adjacency matrix of $\ccalG_c$, constrained to the set $\ccalW = \{ \bbW \in \reals_{+}^{N \times N} : \bbW = \bbW^{\top}, \diag(\bbW) = \mathbf{0} \}$.  Moreover, $\bbE_c$ is the Euclidean-distance matrix obtained from class $c$ signals $\bbY_c:=[\bby_1^{(c)},\ldots,\bby_{T_c}^{(c)}]\in \reals^{N\times {T_c}}$, while $\alpha$, $\beta$, and $\gamma$ are parameters. 

Taking a closer look at the objective function, minimizing $\| \bbW_c \circ \bbE_c\|_{1}$ encourages a graph $\bbW_c$ over which the signals in $\ccalX_c$ are smooth. At the same time, the last term enforces non-smoothness of the signals in the other $C-1$ classes. This composite criterion will thus induce a GFT with better discrimination ability than the baseline $\gamma=0$ case. In fact, the energy of class $c$ signals will be predominantly concentrated in lower frequencies, while the spectral content of the other classes is pushed towards high-pass regions of the spectrum. 
After this training phase, the GFTs of the optimum graphs can be used to extract discriminative features of the test signals via suitable low-pass graph filtering; see~\cite[Section 3.3]{saboksayr20} for details on the classification rule.\vspace{2pt}

\noindent\textbf{EEG emotion recognition.} We apply the discriminative graph learning algorithm for emotion recognition using EEG signals from the DEAP dataset~\cite{deap}. Data were recorded while $32$ subjects watched one-{}minute long music videos. Each participant rated 40 music videos (thus 40 trials), each in terms of the levels of valence, arousal, like/dislike, dominance, and familiarity. We focus on binary (low versus high) valence classification; see~\cite{saboksayr21icassp} for additional details and expanded results. 

\begin{figure}[t]
    \centering
    \begin{minipage}[c]{0.46\linewidth}
    \includegraphics[width=0.6\linewidth]{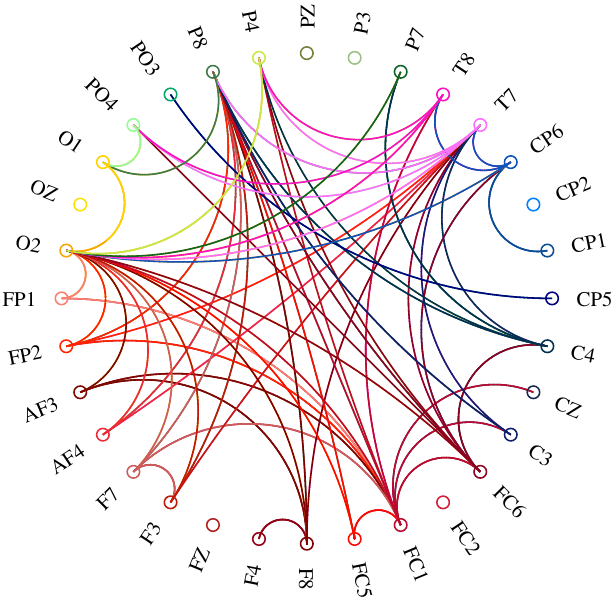}\\
    \centering{\small (a)}
    \end{minipage}
    \begin{minipage}[c]{.17\textwidth}
    \includegraphics[width=\textwidth]{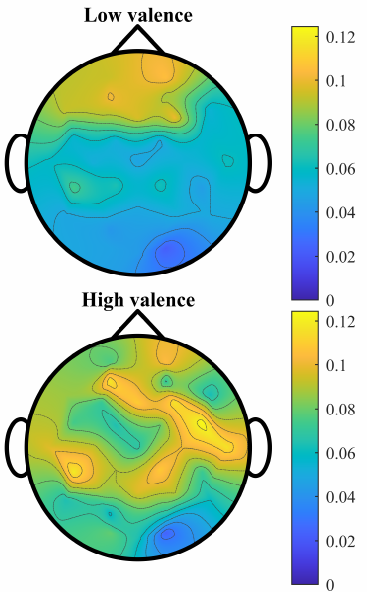}
    \centering{\small (b)}
    \end{minipage}
    \begin{minipage}[c]{.17\textwidth}
    \includegraphics[width=\textwidth]{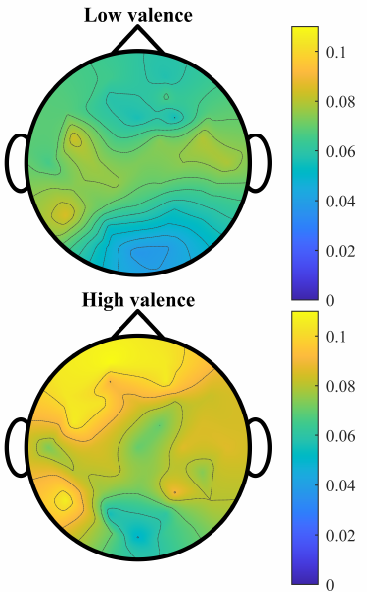}
    \centering{\small (c)}
    \end{minipage}
    \begin{minipage}[c]{.17\textwidth}
    \includegraphics[width=\textwidth]{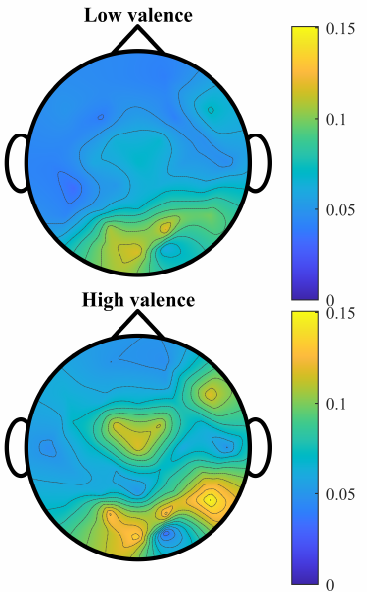}
    \centering{\small (d)}
    \end{minipage}
    \caption{(a)~Significantly different connections between low and high valence with $p\leq 0.002$. Mean of the eigenvector magnitudes corresponding to (b)~low, (c)~mid, and (d)~high frequencies. These results depict distinct discernible patterns between low and high valence, where most of these differences are aligned with the literature; see also~\cite{saboksayr21icassp}.}
    \label{fig:eeg}
\end{figure}

We conduct the classification task in a leave-{}one-{}trial-{}out fashion, where for each subject we use $39$ trials as the training set and test on the one remaining trial. We repeat this $40$ times and report the mean accuracy. 
For the aforementioned discriminative graph learning approach \eqref{E:discriminative}, the reported average classification accuracy across all trials and participants is $92.73\%$, markedly outperforming state-{}of-{}the-{}art methods in valence classification~\cite[Table 1]{saboksayr21icassp}. Specifically, there is a $6\%$ gain over the method in~\cite{kalofolias16}, which can be directly traced to the added discriminative term when $\gamma\neq 0$.

It is also of interest to investigate whether there are useful patterns in the learned class-conditional graphs and the resulting GFT bases. To this end, we study the connections that are significantly different between groups. We also depict the learned eigenvectors corresponding to low, mid, and high frequencies to assess whether our findings are aligned with the literature.
In Fig.~\ref{fig:eeg} (a), we show the significantly ($p<0.002$) different connections between low and high valence obtained using the nonparametric Wilcoxon rank-{}sum test. The original DEAP paper identifies the most discriminative channels~\cite[Table 4]{deap} for valence classification, and almost all of them are also present in Fig.~\ref{fig:eeg} (a). Similar conclusions follow from the eigenvectors in Fig.~\ref{fig:eeg} (b)-(d); see~\cite{saboksayr21icassp} for a detailed analysis. 


\section{Directional, nonlinear, and dynamic interactions}\label{sec:dnd}

So far, the focus has been on learning static and undirected graphs from data. Next, we first consider the identification of directed (di)graphs given nodal time series, which is intimately related to the problem of causal inference.  We then cross the boundary of linear time-invariant network models and outline recent tensor advances for tracking topologies of dynamic graphs, as well as kernel-based methods to account for nonlinear pairwise interactions among vertex processes. Joint inference of the
network topology and processes from partial nodal observations are also outlined.

\subsection{Directed graphs and causality}
\label{ssec:digraphs_causal}

Undirected graphs, like correlation networks, can inform proximity between nodal signal values but cannot inform causality.
Here, we will lift the assumption that adjacency matrices are symmetric and consider estimation of digraphs~\cite{marques20} with the intention of inferring causality from either snapshot or time series observations~\cite{elements}.

To that end, structural equation models (SEMs) comprise a family of statistical methods that model directional (possibly causal) relationships between interacting variables in a complex system. This is pursued through estimation of linear relationships among endogenous as well as exogenous traits. SEMs have been extensively
adopted in economics, psychometrics, social sciences, and genetics, among others; see e.g.,~\cite{kaplan_book}. The appeal of SEMs can be attributed to simplicity and their inherent ability to capture edge directionality in graphs, represented through a (generally) asymmetric adjacency matrix $\bbW\in\reals^{N\times N}$ whose entry $W_{ij}$ is nonzero only if a directed edge connects nodes $i$ and $j$ (pointing from $j$ to $i$). 

SEMs postulate a linear time-invariant network model of the form
\begin{equation}\label{E:SEM}
y_{it}=\sum_{j=1, j\neq i}^N W_{ij}y_{jt}+B_{ii}x_{it}+e_{it}, \: i\in \ccalV \:\: \Rightarrow \bby_t=\bbW\bby_t+\bbB\bbx_t+\bbe_t
\end{equation}
where $\bby_t := [y_{1t},\ldots,y_{Nt}]^\top$ represents a signal of endogenous nodal variables at discrete time $t$ and $\bbx_t :=[x_{1t},\ldots,x_{Nt}]^\top$ is a vector of exogenous influences. The term $\bbW\bby_t$ in \eqref{E:SEM} models network effects, implying $y_{it}$ is a linear combination of the instantaneous values $y_{jt}$ of node $i$'s in-neighbors $j\in\ccalN_i$. The signal $y_{it}$ also depends on $x_{it}$, where $B_{ii}$ captures the level of influence of external sources, thus defined as $\mathbf{B}:=\textrm{diag}(B_{11},\ldots,B_{NN})$. Depending on the context, $\bby_t$ can be thought of as an output signal while $\bbx_t$ corresponds to the excitation or control input. Vector $\bbe_t$ with independent entries accounts for measurement errors and unmodeled dynamics. 


Given i.i.d. snapshot data $\{\bby_t,\bbx_t\}_{t=1}^T$, SEM parameters $\bbW$ and $\bbb:=[B_{11},\ldots,B_{NN}]^\top$ are typically estimated via penalized least squares (LS), for instance by solving~\cite{BazerqueGeneNetworks}
\begin{align}\label{E:SEM_estimation}
&{}\min_{\bbW,\bbb}\left\{\sum_{t=1}^T\|\bby_t-\bbW\bby_t-\bbB\bbx_t\|^2+\alpha\|\bbW\|_1\right\}\\
&{}\textrm{subject to }  \quad\textrm{diag}(\bbW)=\mathbf{0},\quad \bbB=\textrm{diag}(\bbb) \nonumber
\end{align}
where the $\ell_1$-norm penalty promotes sparsity in the adjacency matrix. Both edge sparsity as well as the endogenous inputs play a critical role towards guaranteeing the SEM parameters \eqref{E:SEM} are uniquely identifiable; see also~\cite{giannakis18}.\vspace{2pt}

\noindent\textbf{Characterizing acyclicity for directed acyclic graph (DAG) estimation.} When $\ccalG$ is directed and \emph{acyclic} (DAG), the SEM in \eqref{E:SEM} induces a joint distribution for $\bby=[y_{1},\ldots,y_{N}]^\top$ (we drop the $t$ dependency for convenience), whereby each random variable $y_i$ depends solely on its parents $\textrm{PA}_i = \{j \in \ccalV : W_{ij} \neq 0\}$. These probabilistic graphical models are known as Bayesian networks and have well-appreciated merits for encoding causal relationships within complex systems~\cite{elements}. Recognizing that DAG learning from observational data [i.e., \eqref{E:SEM_estimation} with added acyclicity constraints on $\bbW$] is in general an NP-hard problem, recent efforts have advocated a continuous relaxation approach which offers an efficient means of exploring the space $\mathbb{D}$ of DAGs~\cite{noTears,golem,dagma}. Noteworthy methods advocate an exact acyclicity characterization using nonconvex, smooth functions $\ccalH:\reals^{N\times N}\mapsto \reals$ of the adjacency matrix, whose zero level set is $\mathbb{D}$. One can thus relax the combinatorial constraint $\ccalG(\ccalV,\ccalE,\bbW) \in \mathbb{D}$ by instead enforcing $\ccalH(\bbW)=0$, and tackle the DAG learning problem using standard continuous optimization algorithms \cite{noTears, poly, noFears, dagma,saboksayr2024colide}. The pioneering NOTEARS formulation introduced $\ccalH_{\text{expm}}(\bbW) = \operatorname{trace}(e^{\bbW \circ \bbW}) - N$~\cite{noTears}. Diagonal entries of powers of $\bbW \circ \bbW$ encode information about cycles in $\ccalG$, hence the suitability of the chosen function. Follow-up work suggested a more computationally efficient acyclicity function  $\ccalH_{\text{poly}}(\bbW) = \operatorname{trace}((\bbI + \frac{1}{N} \bbW \circ \bbW)^{N}) - N$~\cite{poly}; see also~\cite{noFears} that studies the general family $\ccalH(\bbW) = \sum_{k=1}^N c_k\operatorname{trace}((\bbW \circ \bbW)^{N})$, $c_k> 0$. Recently, \cite{dagma} proposed the log-determinant acyclicity characterization $\ccalH_{\text{ldet}}(\bbW, s) = N \log(s) - \log(\det(s \bbI - \bbW \circ \bbW))$, $s \in \reals$; outperforming prior relaxation methods in terms of (nonlinear) DAG recovery and efficiency.\vspace{2pt}

\noindent\textbf{Vector autoregressive models for time series data.} While SEMs only capture contemporaneous relationships among the nodal variables (meaning SEMs are memoryless), the class of vector autoregressive models (VARMs) can further account for linear time-lagged (causal) influences; see e.g.,~\cite{varm_group_sparse,timeseries2005book}. Specifically, for given model order $L$ and unknown (typically sparse) evolution matrices $\{\bbW^{(l)}\}_{l=1}^L$, VARMs postulate a multivariate linear dynamical model of the form
\begin{equation}\label{E:SVARM}
\bby_t=\sum_{l=1}^L\bbW^{(l)}\bby_{t-l}+\bbe_t.
\end{equation}
%
Here, a directed edge from vertex $j$ to $i$ is present in $\ccalG$ if at least one of $\{W_{ij}^{(l)}\}_{l=1}^L$ is nonzero (the OR rule). The other common alternative relies on the AND rule, which requires $W_{ij}^{(l)}\neq 0$ for all $l=1,\ldots,L$ to have $(i,j)\in \ccalE$. The AND rule is often explicitly imposed as a constraint during estimation of VARM parameters, through the requirement that all matrices $\bbW^{(l)}$ have a common support. This can be achieved for instance via a group lasso penalty, that promotes sparsity over edgewise coefficients $\bbw_{ij}:=[W_{ij}^{(1)},\ldots,W_{ij}^{(L)}]^\top$ jointly~\cite{varm_group_sparse}. The sparsity assumption is often well-justified due to physical considerations or for the sake of interpretability, but here it is also critical to reliably estimate the graph from limited and noisy data.

The benefits of SEMs and VARMs can be leveraged jointly through so-termed \emph{structural} (S)VARMs, which augment the right-hand-side of \eqref{E:SVARM} with a term $\bbW^{(0)}\bby_t$ to also capture instantaneous relationships among variables as in \eqref{E:SEM}; see also~\cite{giannakis18}. In~\cite{willet_autoregressive}, the inference of the autoregressive parameters and associated network structure is studied within a generalized SVARM framework that includes discrete Poisson and Bernoulli autoregressive processes. SVARMs are also central to popular digraph topology identification approaches based on the principle of Granger causality~\cite{granger}. Said principle is based on the concept of precedence and predictability,
where node $j$'s time series is said to ``Granger-cause'' the time series at node $i$ if knowledge of $\{y_{j,t-l}\}_{l=1}^L$
improves the prediction of $y_{it}$ compared to using only $\{y_{i,t-l}\}_{l=1}^L$. Such form of causal dependence defines the status of a candidate edge from $j$ to $i$, and it can be assessed via judicious hypothesis testing~\cite{timeseries2005book}. 

\subsection{Nonlinear models for topology identification}
\label{ssec:nonlinear}

Network models such as SEMs or SVARMs are linear, and the same is true for most measures of pairwise similarity we encountered in Section \ref{sec:undirected}; notably those based on Pearson or partial correlations. However, in complex systems such as the brain, there is ample evidence that dynamics involve nonlinear interactions between neural regions, and accordingly linear models fall short in capturing such dependencies. 

Recognizing these limitations of linear models, several nonlinear variants of SEMs have emerged; see e.g.,~\cite{joreskog1996nonlinear,wall2000estimation,jiang2010bayesian,lee2003model,harring2012comparison,kelava2014nonlinear,shen2019nonlinear}. A limitation is that these works assume that the graph topology is \emph{known a priori}, and the algorithms developed only estimate the unknown edge weights. On the other hand, several variants of nonlinear GC and VARMs have well-documented merits in unveiling links that often remain undiscovered by traditional linear models; see e.g.,\cite{marinazzo2008kernel,marinazzo2008kernel2,sun2008assessing,lim2015operator}. Linear and nonlinear GC metrics on the other hand entail multiple pairwise tests. These considerations motivate the ensuing approach that jointly identifies edges by leveraging sparse nonlinear SVARMs. 

Consider first linear SVARMs that postulate $y_{jt}$ as a linear combination of instantaneous measurements at the remaining nodes $\{ y_{it} \}_{i \neq j}$, and their time-lagged counterparts $ \{ \{ y_{i(t-\ell)}\}_{i=1}^N \}_{\ell = 1}^{L}$~ \cite{chen2011vector}. Specifically, $y_{jt}$ obeys the model
 \begin{align}
 	y_{jt}= \sum_{i\neq j}W_{ij}^{(0)}y_{it} + \sum_{i=1}^N\sum_{\ell=1}^{L} W_{ij}^{(\ell)} y_{i(t-\ell)} +e_{jt}
	\label{eq:svar:linear}
 \end{align}
where $W_{ij}^{(\ell)}$ for $\ell \neq 0$ captures the \emph{causal} influence of node $i$ on node $j$ over a lag of $\ell$ slots, while $W_{ij}^{(0)}$ encodes the corresponding instantaneous relationship between the two. A link is present from node $i$ to node $j$ either when $W_{ij}^{(0)} \neq 0$, or, when there is a certain $\ell \in \{1,\ldots,L\}$ for which $W_{ij}^{(\ell)} \neq 0$. Order $L$ can be determined via model selection methods such as the Bayesian information~\cite{chen1998speaker}, or, Akaike's criteria~\cite{bozdogan1987model}.

Consider the multivariate nonlinear regression [cf. \eqref{eq:svar:linear}] $\bby_{t}=\bar{\bbf}(\bby_{\backslash jt}, \{\bby_{t-\ell}\}_{\ell=1}^L)+\bbe_t$, with entrywise form
$y_{jt}=\bar{f}_j(\bby_{\backslash jt}, \{\bby_{t-\ell}\}_{\ell=1}^L)+e_{jt},\quad j=1,\ldots, N$.
To circumvent the `curse of dimensionality' in estimating a $[(L+1)N-1]$-variate function, we will confine our multivariate function $\bar{f}_j$ to be separable wrt each of its $(L+1)N-1$ variables. Such a simplification amounts to adopting a generalized additive model~\cite[Chap. 9]{trevor2009elements}, here of the form  $\bar{f}_j(\bby_{\backslash jt}, \{\bby_{t-\ell}\}_{\ell=1}^{L})=\sum_{i\neq j} \bar{f}^{(0)}_{ij}(y_{it}) + \sum_{i=1}^N\sum_{\ell=1}^{L}  \bar{f}^{(\ell)}_{ij}  (y_{i(t-\ell)})$.
With  $\bar{f}_{ij}^{(\ell)}(y):=W_{ij}^{(\ell)} f_{ij}^{(\ell)} (y)$, and postulating that the node $j$ measurement at $t$ depends on instantaneous spatial and time lagged effects, one arrives at [cf. \eqref{eq:svar:linear}]
\begin{align}
\label{eq:svar:nonlinear}
\hspace*{-0.2cm}	y_{jt}=\sum_{i\neq j}W^{(0)}_{ij} f^{(0)}_{ij}(y_{it}) + \sum_{i=1}^N\sum_{\ell=1}^{L} W^{(\ell)}_{ij}   f^{(\ell)}_{ij}  (y_{i(t-\ell)}) +e_{jt}
\end{align}
where similar to \eqref{eq:svar:linear}, $\{W_{ij}^{(\ell)}\}$ specify the lag-adjacency  matrices $\{\bbW^{(\ell)}\}_{\ell=0}^{L}$. 
Rather than the $[(L+1)N-1]$-variate $\bar{f}_j$, \eqref{eq:svar:nonlinear} requires estimating $(L+1)N-1$ \emph{univariate} functions $\{f_{ij}^{(\ell)}\}$. 

The linear SVARM in \eqref{eq:svar:linear} presumes that $\{f_{ij}^{(\ell)}\}$ in \eqref{eq:svar:nonlinear} are linear, but upon resorting to an RKHS model of $\{f_{ij}^{(\ell)}\}$ nonlinear renditions can be accommodated too \cite{shen2017tsp1}. Let each univariate $f_{ij}^{(\ell)}(.)$ in \eqref{eq:svar:nonlinear} belong to the RKHS $\mathcal{H}_i^{(\ell)}:=\{f_{ij}^{(\ell)}|f_{ij}^{(\ell)}(y)=\sum_{t=1}^{\infty} \beta_{ijt}^{(\ell)}\kappa_i^{(\ell)}(y, y_{i(t-\ell)})\}$. 
Considering the measurements at node $j$, and $f_{ij}^{(\ell)}\in \mathcal{H}_i^{(\ell)}$, for $i=1,\ldots, N$ and $\ell=0,1,\ldots, L$, the  regularized LS estimates of these functions are
\begin{align}
	&\{\hat{f}_{ij}^{(\ell)}\}=\arg \min_{\{f_{ij}^{(\ell)}\in \mathcal{H}_i^{(\ell)}\}}  \frac{1}{2}\sum_{t=1}^T\bigg[y_{jt}-\sum_{i\neq j}W^{(0)}_{ij} f^{(0)}_{ij}(y_{it})
	\label{eq:prob:0} \\
	& \hspace*{-0.25cm}- \sum_{i=1}^N\sum_{\ell=1}^{L} W^{(\ell)}_{ij}   f^{(\ell)}_{ij}  (y_{i(t-\ell)}) \bigg]^2+\lambda \sum_{i=1}^N\sum_{\ell=0}^{L} \Omega\left(\|W_{ij}^{(\ell)} f_{ij}^{(\ell)}\|_{\mathcal{H}^{(\ell)}}\right) \nonumber
\end{align}
where the regularizer $\Omega(\bbz)$ can be chosen to effect different attributes, such as sparsity using the $\Omega(\boldsymbol{\zeta})=\|\boldsymbol{\zeta}\|_1$ surrogate of the $\ell_0$-norm \cite{donoho2006most}. Invoking again the representer theorem~\cite[p.\;169]{trevor2009elements}, the optimal
$\hat{f}_{ij}^{(\ell)}(y)=\sum_{t=1}^T \beta_{ijt}^{(\ell)}\kappa_i^{(\ell)}(y,y_{i(t-\ell)})$
can be substituted into \eqref{eq:prob:0}, and with  $\bbbeta_{ij}^{(\ell)}:=[\beta_{ij1}^{(\ell)}, \ldots, \beta_{ijT}^{(\ell)}]^\top$, $\bbalpha_{ij}^{(\ell)}:=W_{ij}^{(\ell)} \bbbeta_{ij}^{(\ell)}$, the functional minimization in \eqref{eq:prob:0} boils down to optimizing over vectors $\{\bbalpha_{ij}^{(\ell)}\}$ to find 
\begin{align}
\label{eq:obj:1}
&\hspace{-0.3cm}\{\hat{\bbalpha}_{ij}^{(\ell)}\}=\arg \hspace*{-0.1cm} \min_{\{\bbalpha_{ij}^{(\ell)}\}} \frac{1}{2}\bigg\|\bby_{j} \hspace*{-0.1cm}  -  \hspace*{-0.1cm} \sum_{i\neq j}\bbK_{i}^{(0)}\bbalpha_{ij}^{(0)} 
\hspace*{-0.1cm}	- \hspace*{-0.1cm} \sum_{i=1}^N\sum_{\ell=1}^L \bbK_{i}^{(\ell)}\bbalpha_{ij}^{(\ell)}  \bigg\|_2^2\nonumber\\
	&\hspace{2.5cm}+\lambda \sum_{i=1}^N\sum_{\ell=0}^L\Omega\bigg(\sqrt{(\bbalpha_{ij}^{(\ell)})^\top\bbK_i^{(\ell)}\bbalpha_{ij}^{(\ell)}}\bigg)
\end{align}
where the $T\times T$ matrices $\{\bbK_i^{(\ell)}\}$ have entries $[\bbK_i^{(\ell)}]_{t,t'}=\kappa_i^{(\ell)}(y_{it}, y_{i(t'-\ell)})$ expressed using kernels $\{\kappa_i^{(\ell )} (\cdot)\}$. The nonzero $W_{ij}^{(\ell)}$ specifying the topology can be found as the solution of  
\eqref{eq:obj:1} using the ADMM; see  e.g.,~\cite{Giannakis2016}. 

Rather than pre-selecting $\{\kappa_i^{(\ell)}\}$, a data-driven MKL alternative applies here too~\cite{shen2017tsp1}. Consider just for notational simplicity that  $\kappa_i^{(\ell)}=\kappa \in\mathcal{K} $, for $\ell=0, 1,\ldots, L$ and $i=1, \ldots, N$ in \eqref{eq:prob:0}; and thus, $\mathcal{H}_i^{(\ell)}=\mathcal{H}^{(\kappa)}$. With $\mathcal{H}_p$ denoting the RKHS induced by $\kappa_p$, the optimal $\{\hat{f}_{ij}^{(\ell)}\}$  is expressible in a separable form as
$\hat{f}_{ij}^{(\ell)}(y): =\sum_{p=1}^Pf_{ij}^{(\ell,p)}(y)$,
where $f_{ij}^{(\ell,p)}$ belongs to RKHS $\mathcal{H}_p$, for $p=1, \ldots, P$ \cite{bazerque2013nonparametric,micchelli2005learning}. Hence, \eqref{eq:prob:0} with data-driven kernel selection reduces to
\small 
\begin{align}
	\label{eq:prob:mkl1}
	&\{\hat{f}_{ij}^{(\ell)}\}=\arg   \min_{\{f_{ij}^{(\ell,p)}\in \mathcal{H}_p\}}
	\frac{1}{2}\sum_{t=1}^T \bigg[y_{jt}-\sum_{i\neq j}\sum_{p=1}^P W^{(0)}_{ij}f^{(0,p)}_{ij}(y_{it})\\
	& \hspace*{-0.1cm} - \hspace*{-0.1cm} \sum_{i=1}^N\sum_{\ell=1}^{L}\sum_{p=1}^P W^{(\ell)}_{ij}   f^{(\ell,p)}_{ij}  (y_{it}) \bigg]^2
\hspace*{-0.1cm} + \hspace*{-0.1cm} ~\lambda~ \sum_{i=1}^N\sum_{\ell=0}^{L}\sum_{p=1}^P\Omega\left( \|W_{ij}^{(\ell)} f_{ij}^{(\ell,p)}\|_{\mathcal{H}_p}\right).\nonumber
\end{align}
\normalsize
As \eqref{eq:prob:mkl1} and \eqref{eq:prob:0} only differ in the extra summation over $P$ kernels, \eqref{eq:prob:mkl1} can also be implemented by an efficient solver~\cite{shen2017tsp1}. 

The kernel-based SVARM outlined here can identify the topology of directed graphs. By simply including linear kernels in the dictionary, it subsumes also linear SVARMs.   It can further account for nonlinear interactions, as well as sparsity and low rank of adjacency matrices, while at the same time it scales well with the number of data and the graph size. In a nutshell, the MKL-based RKHS methodology  offers a principled overarching approach to topology identification.

\subsection{Dynamic networks and multilayer graphs}
\label{ssec:dynamic_tensor}

As data become more complex and heterogeneous, possibly generated in a streaming fashion by nonstationary sources, it is becoming increasingly common to rely on models comprising \textit{multiple} related networks describing the interactions between various entities. While dynamic graphs with time-varying topologies naturally fall within this general class of models, the multiple graphs of interest need not be indexed by time, but possibly instead by different subjects, demographic variables, or sensing modalities. This is for instance the case in neuroscience, where observations for different patients are available and the objective is to estimate their functional brain networks; and in computational genomics where the goal is to identify pairwise interactions between genes when measurements for different tissues of the same patient are available. In order to unveil hidden structures, detect anomalies, and interpret the temporal dynamics of such data, it is essential to understand the relationships between the different entities and how these relationships evolve over time or across different modalities. Joint identification of multiple adjacency matrices can be useful even when interest is only in one of the networks, since joint formulations exploit additional sources of information and, hence, they are likely to result in more accurate topology estimates. Although noticeably less than its single-network counterpart, joint inference of multiple graphs has attracted attention, especially for the case of GMRFs and in the context of dynamic (time-varying) topologies~\cite{giannakis18,Kalofolias2017inference_dynamic,baingana2014cascadesJSTSP14,shen2017tensors,time_varying_graphical_lasso}. 
All of the aforementioned works consider that the multiple graphs $\ccalG_t(\ccalV,\ccalE_t,\bbW_t)$, $t=1,\ldots,T$, share a common vertex set while being allowed to have different edge sets and weights, a structure oftentimes referred to as a multilayer graph or a network of networks. Given the previous motivation, here we extend several of the problem formulations in the previous sections to accommodate (dynamic) multilayer graphs.

To state the joint network topology inference problem in its various renditions, consider a scenario with $T$ different graphs $\ccalG_t(\ccalV,\ccalE_t,\bbW_t)$ defined over a common set $\ccalV$ of nodes, with cardinality $|\ccalV|=N$. This implies the existence of $T$ different adjacency matrices $\{\bbW_t\}_{t=1}^T$ that we want to recover, all represented by $N\times N$ matrices, whose sparsity pattern and nonzero values may be different across $t$. Suppose that for each graph we have access to a set of nodal measurements $\ccalY_t:=\{\bby_t^{(p)}\}_{p=1}^{P_t}$. Equivalently, it will be convenient to represent $\ccalY_t$ through the matrix $\smash{\bbY_t:=[\bby_t^{(1)},\ldots,\bby_t^{(P_t)}]\in\reals^{N\times P_t}}$ containing the $P_t$ signals associated with graph $\ccalG_t$.
\vspace{2pt}

\noindent\textbf{Slowly-varying dynamic networks.} A popular approach to joint inference of multilayer networks assumes that graphs $\ccalG_t$ and $\ccalG_{t-1}$ are \emph{similar} in some application-dependent sense, which we encode as some matrix distance $r(\bbW_t,\bbW_{t-1})$ being small. For instance, this could be well motivated for identification of a sequence of slowly time-varying graphs. In this context, a general formulation entails solving
\begin{equation}\label{E:tv_estimator_general}
\{\bbW^{*}_t\}_{t=1}^T :=\arg\min_{\{\bbW_t\in \ccalW\}_{t=1}^T }\left\{\sum_{t=1}^T\Phi_t(\bbW_t,\bbY_t)+\alpha \sum_{t=2}^Tr(\bbW_t,\bbW_{t-1})\right\}
\end{equation}
where $\Phi_t$ is an often convex objective function stemming from the adopted topology identification criterion, and $\alpha>0$ is a regularization parameter. In~\cite{time_varying_graphical_lasso}, the so-termed time-varying graphical lasso estimator was proposed to identify a collection of GMRFs encoded in the precision matrices $\bbW_t=\bbTheta_t:=\bbSigma_t^{-1}$.  Therein,  $\Phi_t(\bbW_t,\bbY_t)=-\log\det\bbTheta_t +\textrm{trace}(\hbSigma_t \bbTheta_t)+\lambda\|\bbTheta_t\|_1$ corresponds to the penalized global likelihood for $\bby_t\sim\textrm{Normal}(\mathbf{0},\bbSigma_t)$ we encountered in \eqref{E:glasso}. A comprehensive list of distance functions $r$ was also introduced to encode different network evolutionary patterns including smooth and abrupt topology transitions; see~\cite{time_varying_graphical_lasso} for further details. From an algorithmic standpoint, an ADMM solver is adopted to tackle \eqref{E:tv_estimator_general} efficiently.

Recalling the discussion in Section \ref{ssec:smooth}, a general estimator for learning slowly time-varying graphs over which signals in $\ccalY_t$ exhibit smooth variations was developed in~\cite{Kalofolias2017inference_dynamic}. Let $\bbW_t$ denote the adjacency matrix of $\ccalG_t$ and recall the Euclidean-distance matrix $\bbE_t$ defined in Section \ref{ssec:smooth}. The idea in~\cite{Kalofolias2017inference_dynamic} is to set $\Phi_t(\bbW_t,\bbY_t)=\|\bbW_t\circ\bbE_t\|_1-\alpha\mathbf{1}^\top\log(\bbW_t\mathbf{1})+\frac{\beta}{2}\|\bbW_t\|_F^2$ in \eqref{E:tv_estimator_general} along with $r(\bbW_t,\bbW_{t-1})=\|\bbW_t-\bbW_{t-1}\|_F^2$, and rely on the primal-dual algorithm of~\cite{kalofolias16} to tackle the resulting separable optimization problem. Through a different choice of $\Phi_t(\bbW_t,\bbY_t)$, the framework therein can also accommodate a time-varying counterpart of the model in~\cite{DongLaplacianLearning}.

On a related note, network topology identification from temporal traces of infection events has emerged as an active area of research, which is relevant to epidemic processes, propagation of viral news events between blogs, 
or acquisition of new buying habits by consumer groups. It has been observed in these settings that information often spreads in
cascades by following implicit links between nodes in a time-varying graph $\ccalG_t$. Reasoning that infection times depend on both topological (endogenous)
and external (exogenous) influences, a \emph{dynamic} SEM-based scheme was proposed
in~\cite{baingana2014cascadesJSTSP14} for cascade modeling. With $c=1,\ldots,C$ indexing cascades, similar to \eqref{E:SEM} one can model topological influences as linear combinations of infection times $y_{it}^{(c)}$ of other nodes in the network, whose weights
correspond to entries in a time-varying asymmetric adjacency matrix $\bbW_t$. External
influences $x_{i}^{(c)}$ such as those due to on-site reporting in news propagation contexts
are useful for model identifiability, and they are taken to be time invariant for simplicity. 
It is assumed that the networks $\ccalG_t$ vary slowly with time, facilitating adaptive SEM parameter estimation by minimizing a
sparsity-promoting exponentially-weighted LS criterion~\cite{baingana2014cascadesJSTSP14}
\begin{align}\label{E:dynamic_SEM_estimation}
&{}\min_{\bbW_t,\bbb_t}\left\{\sum_{c=1}^C\sum_{t=1}^T\gamma^{T-t}\|\bby_t^{(c)}-\bbW_t\bby_t^{(c)}-\bbB_t\bbx^{(c)}\|^2+\alpha_t\|\bbW_t\|_1\right\}\\
&{}\textrm{subject to }  \quad\textrm{diag}(\bbW_t)=\mathbf{0},\quad\bbB_t=\textrm{diag}(\bbb_t),\: t=1,\ldots,T\nonumber
\end{align}
where $\gamma\in (0, 1]$ is the forgetting factor. 
With $\gamma < 1$, past data are exponentially discarded to track  dynamic network topologies. Related likelihood-based approaches have been advocated to identify traces of network diffusion~\cite{gomez}, as well as tensor-based topology identification for dynamic SEMs that can account for (abruptly) switching topological states representing the layers of the graph~\cite{shen2017tensors}, as outlined next.\vspace{2pt}

\noindent \textbf{Tensor-based methods for piecewise-constant graph sequences.} Here we postulate that the graph has a \emph{piecewise-constant} topology, modeled by a sequence of unknown adjacency matrices $\{ \bbW_m \in \mathbb{R}^{N\times N}, \; t \in [\tau_m, \tau_{m+1} - 1] \}_{m=1}^M$, over $M$ time segments. The $(i,j)$-th entry  $[\bbW_m]_{ij}=W_{ij}^m$ {is nonzero only if a directed edge links node $j$ to $i$ over time segment $m$.} The observations obey time-varying SEMs; that is, $y_{jt}=\sum_{i\neq j}W_{ij}^my_{it} +B_{jj}^m x_{jt} +e_{jt}$ for $t \in [\tau_m,\tau_{m+1} - 1]$
per segment $m = 1,\dots, M$, with $e_{jt}$ capturing unmodeled dynamics, while coefficients $\{ W_{ij}^m \}$ and $\{ B_{jj}^m \}$ are unknown.
The noise-free matrix-vector SEM is then ${\bby}_t=\bbW_m{\bby}_t+\bbB_m\bbx_t$, where $[\bbW_m]_{ij} = W_{ij}^m$ and $\bbB_m := \text{diag}(B_{11}^m, \dots, B_{NN}^m)$. 
Suppose also that the exogenous inputs $\{ \bbx_t^{(m)} \}$ are \emph{piecewise-stationary} {over segments $t\in[\tau_{m}, \tau_{m+1} - 1], m=1,\dots,M+1$}, {each with} a fixed correlation matrix $\{ \bbR_m^x := \mathbb{E}[ \bbx_t^{(m)} (\bbx_t^{(m)})^{\top} ]\}_{m=1}^M$. Under these conditions, an online algorithm can be developed for tracking $\{ \bbW_m, \bbB_m \}_{m=1}^M$ using measured endogenous variables, and the correlation matrices $\{ \bbR_m^x\}_{m=1}^M$~\cite{shen2016tensor,shen2017tensors}.

To this end, let $\boldsymbol{\mathcal{A}}_m:=(\bbI-\bbW_m)^{-1}\bbB_m$, and consider a tensor $\underbar{\bbR}^y$ with its $m$-th slice $\bbR_m^y 	= \boldsymbol{\mathcal{A}}_m \bbR_m^x \boldsymbol{\mathcal{A}}_m^{\top}, \quad t \in [\tau_{m}, \tau_{m+1} - 1]$ sequentially appended at $t=\tau_{m+1}$, for $m = 1, \dots, M$. If $\mathbb{E} \{ x_{it} x_{jt} \} = 0, \forall\: i \neq j$, the $m$th slice can be expressed as a weighted sum of rank-one matrices
{\color{black}
\begin{align}\label{eq:R}
	\bbR_m^y = \boldsymbol{\mathcal{A}}_m \text{diag}(\boldsymbol{\rho}^x_m) \boldsymbol{\mathcal{A}}_m^\top \end{align}
where $\boldsymbol{\rho}^x_m := [\rho^x_{m1} \ldots \rho^x_{mN}]^\top$, with $\rho^x_{mi} :=\mathbb{E}(x_{it}^2)$, for $t\in  [\tau_{m}, \tau_{m+1} - 1]$; see also Fig. 3 in 
\cite{shen2017tensors}. }

Allowing $\underbar{\bbR}^y$ to grow sequentially along one mode is well motivated for real-time operation, where data may be acquired in a streaming manner. In this case, unveiling the evolving topology calls for approaches capable of tracking tensor factors $\boldsymbol{\mathcal{A}}_m$. Given the tensor  $\underbar{\bbR}^y$, and possibly $\bbR_x$, algorithms for tracking dynamic tensor factors, e.g., PARAFAC via recursive least-squares tracking (PARAFAC-RLST), can be employed; see also ~\cite{nion2009adaptive,shen2017tensors,shen2016tensor} for details. Once $\widehat{\boldsymbol{\mathcal{A}}}_m$ is obtained,
$\bbW_m$ can be estimated on the fly as
$
\hat{\bbW}_m = \bbI-\left(\text{diag}(\widehat{\boldsymbol{\mathcal{A}}}^{-1}_m)\right)^{-1}\widehat{\boldsymbol{\mathcal{A}}}^{-1}_m$ \cite{shen2017tensors}.

Tensor-based topology identification along these lines applies to both dynamic and static graphs, so long as 
(even a subset of) second-order statistics of the exogenous inputs are available, and change across segments; 
see~\cite{shen2017tensors} and~\cite{shen2017asilomar}, where identifiability is studied under low-rank and sparsity constraints on the adjacency matrix. Thus, piece-wise input stationary correlations play a role analogous to multiple layers, time-lagged and nonlinear terms in SVARMs, or, the exogenous variables themselves in linear SEMs - what can be critical for identifability when inputs cannot be available (e.g., due to privacy concerns), but their statistics can be measured.

Even though CPD based SEM (CPSEM) can identify the network structure without exact information about the per-node exogenous input, it faces extra challenges.\\ 
(c1) The approach in \cite{shen2016tensor} relies on the ALS algorithm to perform the CPD for estimating ${\cal A}$, which leads to serious scalability issues. The latent factor ${\cal A}$ in CPSEM has size $N\times N$, and the tensor rank is $N$. Hence, the key ALS operation involving the {\it matricized tensor times Khatri-Rao product} (MTTKRP) costs $O(N^4M)$ flops; see \cite{sidiropoulos2016tensor}. Note that $N$ can be millions in a complex network, and MTTKRP is carried out three times in {\it each iteration} of ALS for this particular problem;\\
(c2) Matrix $\boldsymbol{\mathcal{A}}$ does not capture the structure of $\bbW$, e.g., sparsity, since it is obtained by first inverting $\bbW$; and\\ 
(c3) Even though exact information of $\bbx_t$ is no longer needed for the tensor-based algorithm to identify the topology, it is still necessary to know the second-order statistics $\bbR_x$~\cite{shen2016tensor}, which may not be available in certain applications. 

Challenges (c1)-(c3) can be addressed via joint diagonalization (JD) of the tensor slices. To this end, consider rewriting SEM [cf.\eqref{E:SEM}] as
\begin{align}
\label{eq:jd1}
	(\bbI-\bbW)\bby_t=\bbB \bbx_t
\end{align}
and with $\bbH:=\bbI-\bbW$, write the \emph{per-segment} correlation matrix 
$\bbR_m^y := \mathbb{E}\{ \bby_t \bby_t^{\top} \}$,  as 
\begin{align}
\label{eq:Rx}
	\bbH \bbR_m^y\bbH^\top =\bbB\bbR_m^x\bbB^\top
\end{align}
where $\bbB$ and $\bbR_m^x$ are unknown diagonal matrices as per (as1). Clearly, \eqref{eq:Rx} implies that $\bbR_m^y,~ m=1,\ldots, M$ are jointly diagonalizable by $\bbH$. In this noise-free setup, the latter yields 
$
\bbH \bbR_m^y\bbH^\top=\text{diag}(\bbH \bbR_m^y\bbH^\top),~ m=1, \ldots, M.
$
With $h_{ij}$ denoting the $(i, j)$th entry of $\bbH$, it is easy to see that $h_{ij}$ satisfies
$
			h_{ii}=1,\:
			h_{ij}=-W_{ij},\:  i\neq j
$,
which suggests identifying $\bbH$ by solving 
\begin{align} 
\label{prob:JD1}
	&\min_{\bbH} \sum_{m=1}^M\|\bbH \bbR_m^y\bbH^\top-\text{diag}(\bbH \bbR_m^y\bbH^\top)\|_F^2\nonumber\\
	& \textrm{subject to}~~ h_{ii} = 1.
\end{align}
Although intuitively pleasing, further reflection on \eqref{prob:JD1} reveals that if $\bbH^*$ is an optimal solution to \eqref{prob:JD1}, then any $\bar{\bbH}=\bbH^*\bbPi\bbLambda$ is also an optimal solution, where $\bbPi$ denotes a permutation matrix, and $\bbLambda$ is a nonsingular diagonal scaling matrix. This is because column permutation and scaling ambiguities are intrinsic to JD \cite{belouchrani1997blind} (as in tensor and matrix decompositions), and cannot be removed without extra information.

Nonetheless, permutation and scaling ambiguities are not tolerable in network topology identification, since they make isomorphic graphs indistinguishable. In order to resolve the permutation ambiguity,  \cite{shen2016tensor} relies on additional information about the second-order statistics of the exogenous variables, that is $\bbR^x$. This idea can also be used in the JD-based approach. However, there is an alternative to remove the aforementioned  ambiguities. This alternative utilizes the notion of {\it anchor nodes} to pin down the final estimate of the network topology, where anchor here refers to a node whose connectivity patterns are known {\it a priori} (cf. Examples 1-2). 

Let us further assume that a subset of $[\bbW]_{ij}$ entries with $(i,j)\in\Omega$ is known; see also~\cite{rasoul20}. Incorporating the known entries as constraints in \eqref{prob:JD1} yields the constrained LS problem
\begin{align}
	\label{prob:JD:ls}
	&\min_{\bbH} \sum_{m=1}^M\|\bbH \bbR_m^y\bbH^\top-\text{diag}(\bbH \bbR_m^y\bbH^\top)\|_F^2\nonumber\\
	& \textrm{ subject to }~~ h_{ii} = 1,\qquad h_{ij}=-W_{ij}~~ (i,j)\in\Omega.
\end{align}
Through this formulation, it is possible to estimate $\bbH$, and then the adjacency matrix $\bbW$.
Note that \eqref{prob:JD:ls} is not a standard JD formulation because it includes special constraints. Nevertheless, it can be tackled following ideas from classic JD algorithms, e.g., via block coordinate descent (BCD). {Note that BCD converges to a critical point of the optimization problem under certain conditions---e.g., when the block subproblems are strictly convex; see \cite{xu2013block,razaviyayn2013unified,shen2020topology-jd}}.

\subsection{Joint inference of graph topologies and signals}\label{ssec:joint_topo_signal}

Most approaches to topology identification typically
assume that the data or process over the network is observed at all nodes.
However, application-specific constraints may prevent acquiring
network-wide observations. Alleviating the limited flexibility of existing approaches, this section advocates structural models for
processes over graphs, and develops algorithms for joint identification of the network topology and inference of processes evolving over networks from partial nodal observations.

Suppose we have available $M_{t}$ noisy samples of the $t$-th observation vector
\begin{align}
z_{m,t}=y_{n_{m},t}+e_{m,t}, \quad m=1,...,M_{t}
\label{eq:observation_scalar}
\end{align}
where $\mathcal{M}_{t}:=\{n_{1},...,n_{M_{t}}\}$ contains the indices of the sampled vertices, and $e_{m,t}$ models the observation error. With $\mathbf{z}_{t}:=[z_{1,t},...,z_{M_{t},t}]^{\top}$ and $\boldsymbol{e}_{t}:=[e_{1,t},...,e_{M_{t},t}]^{\top}$, the limited observation model is
\begin{align}
\mathbf{z}_{t}=\mathbf{M}_{t}\mathbf{y}_{t}+\boldsymbol{e}_{t}, \quad t=1,...,T
\label{eq:observation_matrix}
\end{align}
where $\mathbf{M}_{t}$ is an $M_{t}\times N$ matrix with entries $\{(m,n_{m})\}_{m=1}^{M_{t}}$ set to one, and the rest set to zero. The broad goal is the \emph{joint inference of the latent network topology and signals evolving over graphs (JISG)} from noisy observations of the latter at subsets of nodes.

Given $\{\mathbf{z}_{t}\}_{t=1}^{T}$ in \eqref{eq:observation_matrix}, we advocate the following regularized LS problem
\begin{align}
\min_{\mathbf{W}\in\mathcal{A},\{\mathbf{y}_{t}\}_{t=1}^{T}}\mathcal{F}(\mathbf{W},\{\mathbf{y}_{t}\}_{t=1}^{T})&:=\sum_{t=1}^{T}||\mathbf{y}_{t}-\mathbf{W}\mathbf{y}_{t}||_{2}^{2} 
+\sum_{t=1}^{T}\frac{\mu}{M_{t}}||\mathbf{z}_{t}-\mathbf{M}_{t}\mathbf{y}_{t}||_{2}^{2}+\rho_{e}(\mathbf{W})
\label{eq:jisg_objective}
\end{align}
where $\rho_{e}(\mathbf{W}):=2\lambda_{1}||\mathbf{W}||_{1}+\lambda_{2}||\mathbf{W}||_{F}^{2}$ is the elastic net penalty~\cite{trevor2009elements}. This nonconvex problem is solved iteratively using a block coordinate descent (BCD) iteration.

Given $\hat{\mathbf{W}}$, the estimates $\{\hat{\mathbf{y}}_{t}\}_{t=1}^{T}$ are found by solving
\begin{align}
\min_{\mathbf{y}_{t}|_{t=1}^{T}}\sum_{t=1}^{T}||\mathbf{y}_{t}-\hat{\mathbf{W}}\mathbf{y}_{t}||_{2}^{2}+\sum_{t=1}^{T}\frac{\mu}{M_{t}}||\mathbf{z}_{t}-\mathbf{M}_{t}\mathbf{y}_{t}||_{2}^{2}
\label{eq:y_objective_decoupled}
\end{align}
which conveniently decouples across $t$ as
\begin{align}
\min_{\mathbf{y}_{t}}g(\mathbf{y}_{t}):=\frac{M_{t}}{\mu}||(\mathbf{I}_{N}-\hat{\mathbf{W}})\mathbf{y}_{t}||_{2}^{2}+||\mathbf{z}_{t}-\mathbf{M}_{t}\mathbf{y}_{t}||_{2}^{2}.
\label{eq:y_objective_t}
\end{align}
This is solved via gradient descent (GD) iterations.
With $\{\hat{\mathbf{y}}_{t}\}_{t=1}^{T}$ available, $\hat{\mathbf{W}}$ is found by solving the following strongly convex problem
\begin{align}
\min_{\mathbf{W}\in\mathcal{A}}\sum_{t=1}^{T}||\hat{\mathbf{y}}_{t}-\mathbf{W}\hat{\mathbf{y}}_{t}||_{2}^{2}+\lambda_{1}||\mathbf{W}||_{1}+\lambda_{2}||\mathbf{W}||_{F}^{2}.
\label{eq:A_objective}
\end{align}
The resultant ADMM solver is detailed in \cite{ioannidis2019semi}  for the joint inference of signals and graph topologies over time (JISGoT).

\begin{figure*}[tpb!]
\begin{minipage}[b]{.48\textwidth}
\centering
\includegraphics[height=3.2cm]{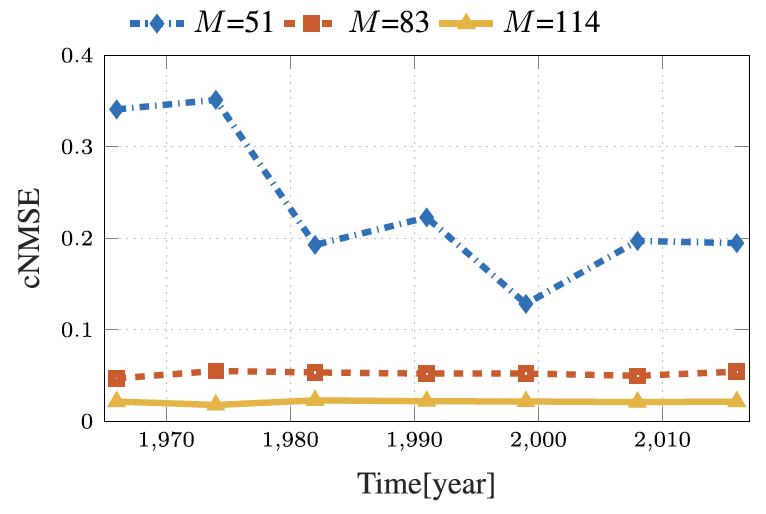}
\centerline{(a)}
\end{minipage}
\begin{minipage}[b]{.48\textwidth}
\centering
\includegraphics[height=3.2cm]{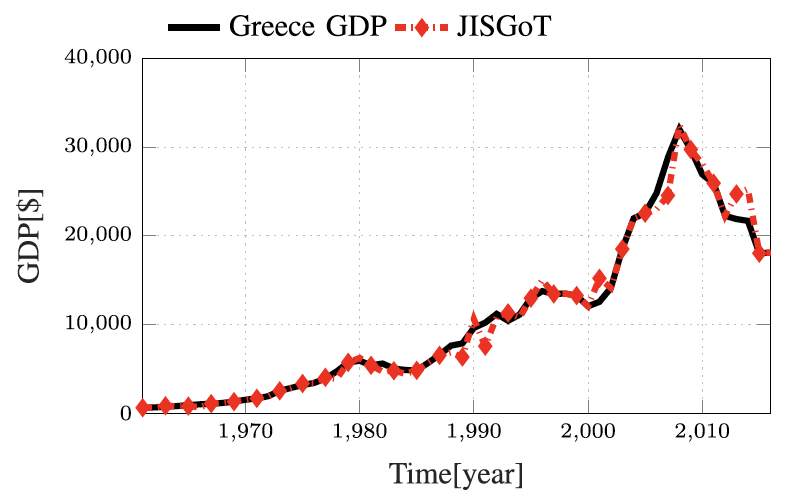}
\centerline{(b)}
\end{minipage}
 \caption{Performance evaluation on GDP prediction: a) cNMSE for GDP estimates; and b) Greece GDP along with JISG estimate; see also \cite{ioannidis2019semi}.} 
 \label{fig:gdp}
\end{figure*}

To evaluate the performance of JISGoT, an experiment is carried over the gross domestic product (GDP) dataset \cite{gdp}, which comprises GDP per capita for $N = 127$ countries over the years $1960-2016$. The process $y_{n,t}$ now denotes the GDP reported at the $n$-th country and $t$-th year for $t = 1960,..., 2016$. The (c)NMSE) is defined as 
$
\mathrm{cNMSE}(T)\ \triangleq\ 
{\sum_{t=1}^{T} \left\| \mathbf{y}_t - \hat{\mathbf{y}}_{t|t} \right\|_2^{2}}/
     {\sum_{t=1}^{T} \left\| \mathbf{y}_t \right\|_2^{2}} \, .
$

Fig. \ref{fig:gdp} (a) depicts the cNMSE of the joint approach for different $M$ values. The semi-blind estimator
unveils the latent connections among countries, while it reconstructs the GDP with $cNMSE = 0.05$ when $60\%$ samples are
available. 
Fig. \ref{fig:gdp} (b) depicts the true values, along with the GDP estimates
of Greece for $M = 89$, which corroborates the effectiveness of JISGoT in predicting the GDP evolution and hence facilitating economic policy planning.

\section{Concluding summary}\label{sec:conclusions}

This chapter surveyed approaches to identifying latent network connectivity and learning signals over graphs, which are prominent problems towards understanding the behavior of complex interconnected systems. Fairly mature statistical approaches are surveyed first, where correlation analysis takes center stage along with its connections to covariance selection, high-dimensional regression for learning Gaussian graphical models, and methods advocating smooth signal assumptions. Reproducing-kernel Hilbert space nonlinear models of pairwise interactions are also discussed, as well as extensions to digraphs and their relation to causal inference. Moreover, topology identification and inference over dynamic graphs were considered as well.

\section*{Acknowledgments}

This work was supported in part by the National Science Foundation under awards CCF-1750428, ECCS-1809356, CCF-1934962, ECCS-2231036, SWIFT-2128593, ECCS-2126052, MoDL-2212318, IMR-2220292, CIF-2312549, ECCS-2412484, ECCS-2442964, and GEO CI-2425748.

\bibliographystyle{vancouver}
%
\bibliography{references.bib,bib4proceed2018ieee}





\Backmatter
\end{document}